\documentclass[intlimits,twoside,a4paper]{article}
\usepackage{amsmath,amssymb}
\usepackage[T2A]{fontenc}
\usepackage[cp1251]{inputenc}
\usepackage[eqsecnum]{cmpj2}

\issue{2016}{19}{3}{33004}
\doinumber{10.5488/CMP.19.33004}

\title{Spectral properties of four-time fermionic Green's functions}
\author{A.M. Shvaika}

\address{Institute for Condensed Matter Physics of the National  Academy of Sciences of Ukraine, \\ 1~Svientsitskii~St., 79011 Lviv, Ukraine}

\DeclareMathOperator{\Tr}{Tr}
\newcommand{\rmd}{\rd}
\newcommand{\rme}{\re}
\newcommand{\rmi}{\ri}
\newcommand{\eref}[1]{(\ref{#1})}

\allowdisplaybreaks

\authorcopyright{A.M.~Shvaika, 2016}
\date{Received April 27, 2016}

\begin{document}
	
	\maketitle
	
	\begin{abstract}
		The spectral relations for the four-time fermionic Green's functions are derived in the most general case. The terms which correspond to the zero-frequency anomalies, known before only for the bosonic Green's functions, are separated and their connection with the second cumulants of the Boltzmann distribution function is elucidated. The high-frequency expansions of the four-time fermionic Green's functions are provided for different directions in the frequency space.
		\pacs 05.30.-d
	\end{abstract}

\section{Introduction}

One of the main tasks of the quantum many-body theory is, on the one hand, to calculate the observable quantities that could be measured directly by experiment, and, on the other hand, to provide connections between the measured quantities and microscopic properties of a system. It was first noticed by Kubo \cite{kubo:570} that linear transport coefficients are expressed in terms of the Fourier transforms of appropriate correlation functions that relate by spectral relations to the two-time Green's functions. Since then, the Green's function method has been admitted and extensively developed \cite{bogolyubov:589,zubarev:320,bonch-bruevich:book}.

In his seminal article Kubo \cite{kubo:570} also pointed out the difference between the isothermal and adiabatic (isolated \cite{wilcox:624}) response of the many-body system and its connection with the ergodic properties of a system. On the other hand, later on it was noticed by Stevens and Toombs \cite{stevens:1307} that spectral relations should be completed by a special treatment of an additional contribution at zero frequency connected with the presence of conserved quantities \cite{suzuki:882,suzuki:277}.

In the Green's function formalism, the issue of ergodicity appears as a difficulty in the determination of the zero-frequency bosonic propagators \cite{stevens:1307,suzuki:882,suzuki:277,fernandez:505,callen:505,lucas:503,morita:1030,kwok:1196,ramos:441}. It states that the Fourier transform of the Green's function contains two terms:
\begin{equation}
G_{AB}(z) = \widetilde{G}_{AB}(z) - C_{AB} \delta(z),
\end{equation}
where the first ergodic (Kubo) contribution $\widetilde{G}_{AB}(z)$ is defined by the one-particle bosonic or fermionic density of states
\begin{equation}\label{eq:dos}
\rho_{AB}(\tilde{\omega})=\frac{1}{Z} \sum_{jf} A_{jf} B_{fj}
\left(\rme^{-\beta\varepsilon_j}\mp \rme^{-\beta\varepsilon_f}\right)\delta(\tilde{\omega}-\varepsilon_{fj})
\end{equation} through the spectral relation
\begin{equation}
\widetilde{G}_{AB}(z) = \int\limits_{-\infty}^{+\infty} \rmd \tilde{\omega} \,\frac{\rho_{AB}(\tilde{\omega})}{z-\tilde{\omega}}\,,
\label{eq:K2erg}
\end{equation}
and the second nonergodic term represents the zero-frequency anomaly with $C_{AB}=0$ for fermionic functions and $C_{AB}\neq0$ for a bosonic one.
Here, the upper and lower signs correspond to bosonic and fermionic functions, respectively,
\begin{equation}
A_{jl}=\langle j|\hat{A}|l\rangle
\end{equation}
are matrix elements of operator $\hat{A}$ between the many-body states with energy difference
\begin{equation}
\varepsilon_{jl}=\varepsilon_{j}-\varepsilon_{l}\,,
\end{equation}
and
\begin{equation}
Z=\Tr \rme^{-\beta H} = \sum_{j} \rme^{-\beta\varepsilon_j}
\end{equation}
is partition function.

For the Matsubara Green's function, the complex argument $z$ is equal to the bosonic or fermionic Matsubara frequencies $z=\rmi\omega_n$, $\delta(z)=\beta\Delta(\rmi\omega_n)$ with $\Delta(z=0)=1$ and $0$ in other cases, and for bosonic functions we have
\begin{equation}\label{eq:K2anom}
C_{AB} = \frac{1}{Z} \sum_{\substack{{jf} \cr {\varepsilon_j=\varepsilon_f}}} \rme^{-\beta\varepsilon_j} A_{jf} B_{fj}\,,
\end{equation}
where the summation is only over the many-body states with equal energies $\varepsilon_j=\varepsilon_f$ including nonergodic contributions and contributions from the conserved quantities.

For the retarded and advanced Green's function one should replace $z$ by $\omega\pm \rmi0^{+}$, respectively, and put $\delta(z)=\delta(\omega)$. Now, the quantity $C_{AB}$ is not well defined and different tricks are used for its calculation \cite{suzuki:882,suzuki:277,fernandez:505,lucas:503,morita:1030,kwok:1196,ramos:441,frobrich:014410,mancini:37,mancini:537}.

Nevertheless, even now many textbooks on the quantum statistics and many-body theory do not provide a complete discussion of the spectral relations and special treatment of the zero-frequency anomalies. Moreover, since for the two-time Green's functions such anomalies appear only for a bosonic one, no one has even tried to address the problem of zero-frequency anomalies for multi-time fermionic Green's functions.

The Kubo's transport theory \cite{kubo:570} is not limited to the linear phenomena and provides results to the arbitrary order of external perturbation. The resulting multi-time correlation and Green's functions represent different nonlinear transport phenomena and resonances \cite{tyablikov:102,tanaka:388,harper:1613}. Besides, multi-time functions also appear as puzzles in different orders of perturbation theories for many-body systems, e.g., the four-time two-particle function enters the Schwinger-Dyson equation for the one-particle function \cite{abrikosov:book}. Moreover, cross-sections of the inelastic scattering processes can be expressed in terms of multi-time correlation functions too \cite{yue:4578,nozieres:3099}, e.g., for the electronic inelastic light (Raman) scattering the nonresonant, mixed, and resonant responses \cite{shastry:1068,devereaux:175,ament:705} are connected with the two-time, three-time, and four-time Matsubara Green's functions \cite{shvaika:137402,shvaika:045120,pakhira:125103}, respectively, and can be rewritten in terms of multi-time correlation functions.

Spectral relations for multi-time, i.e., three-time Green's functions of Kubo type, were originally introduced by Bonch-Bruevich \cite{bonch-bruevich:529,bonch-bruevich:book} before the zero-frequency anomaly problem was noticed. Spectral relations for three-time bosonic Matsubara Green's functions taking into account zero-frequency anomalies were considered by Shvaika \cite{shvaika:447} and there were obtained solutions of the reverse problem, i.e., finding of spectral densities from the known Green's functions.

In this article we consider spectral relations for the four-time fermionic Matsubara Green's functions with special emphasis on zero-frequency anomalies. It will be shown that despite the obvious statement that there are no zero-frequency anomalies for separate fermionic frequencies, they could exist for the sums of two. In the next section we introduce four-time correlation functions and spectral densities and separate the terms with different time and frequency dependences. In section~\ref{sec:4Matsu} we consider the four-time Matsubara Green's functions and show how the zero-frequency anomalies enter and modify the spectral relations. The connection with generalized cumulants will be considered as well. We provide the high-frequency asymptotics in section~\ref{sec:highfrec} and in the last section we conclude.

\section{Four-time correlation functions and spectral densities}

First of all, we should introduce the four-time correlation functions. They can be defined in a usual way as follows:
\begin{equation}
  K_{ABCD}(t_1,t_2,t_3,t_4)=\langle \hat{A}(t_1) \hat{B}(t_2) \hat{C}(t_3) \hat{D}(t_4)\rangle,
\end{equation}
where operators $\hat{A}$, $\hat{B}$, $\hat{C}$, and $\hat{D}$ are of the fermionic type, e.g., ordinary creation and annihilation operators or operators with a more complex commutation relations like the Hubbard one, and
\begin{equation}
\langle\ldots\rangle = \frac{1}{Z} \Tr \rme^{-\beta H} (\ldots)
\end{equation}
is a thermodynamical averaging.
Here, we consider only the case of equilibrium many-body systems for which correlation functions are time-shift invariant
\begin{equation}\label{timeshift}
  K_{ABCD}(t_1,t_2,t_3,t_4)=K_{ABCD}(t_1-t,t_2-t,t_3-t,t_4-t).
\end{equation}

Spectral density is defined as its Fourier transform
\begin{align}
  I_{ABCD}(\omega_1,\omega_2,\omega_3,\omega_4) &=
  \frac{1}{(2\pi)^3}
        \!\!\int\limits_{-\infty}^{+\infty}\!\! \rmd (t_1-t_4)
        \!\!\int\limits_{-\infty}^{+\infty}\!\! \rmd (t_2-t_4)
        \!\!\int\limits_{-\infty}^{+\infty}\!\! \rmd (t_3-t_4)
\nonumber\\
  &\times K_{ABCD}(t_1,t_2,t_3,t_4)
            \rme^ {\rmi(\omega_1 t_1 + \omega_2 t_2 + \omega_3 t_3 + \omega_4 t_4)}
  \Delta(\omega_1+\omega_2+\omega_3+\omega_4).
\label{SD_4t0}
\end{align}
Here, symbol $\Delta(\omega_1+\omega_2+\omega_3+\omega_4)$ represents the conservation of the total energy (frequency)
\begin{equation}\label{freqconserv}
\omega_1+\omega_2+\omega_3+\omega_4=0,
\end{equation}
which follows from the time-shift invariance \eref{timeshift}. \textit{Below, in all equations, we shall keep all four frequencies in order to obtain simple rules for constructing of different contributions, but one should keep in mind that according to \eref{freqconserv} only three of them are independent}.

In the case of fermionic operators, spectral density \eref{SD_4t} includes four different contributions
\begin{align}
I_{ABCD}(\omega_1,\omega_2,\omega_3,\omega_4) &= \Delta(\omega_1+\omega_2+\omega_3+\omega_4)\bigl[\tilde{I}_{ABCD}(\omega_1,\omega_2,\omega_3,\omega_4)
\nonumber\\
 &+ \delta(\omega_1+\omega_2) \bar{I}_{\overline{AB}\;\overline{CD}}(\omega_1,-\omega_1,\omega_3,-\omega_3)
 + \delta(\omega_2+\omega_3) \bar{I}_{\overline{A}\;\overline{BC}\;\overline{D}}(\omega_1,-\omega_3,\omega_3,-\omega_1)
\nonumber\\
 &+\delta(\omega_1+\omega_2)\delta(\omega_2+\omega_3) \bar{\bar{I}}_{\overline{ABCD}}(\omega_1,-\omega_1,\omega_1,-\omega_1)\bigr]
\label{SD_4t}
\end{align}
with different frequency dependences
\begin{align}
 \tilde I_{ABCD}(\omega_1,\omega_2,\omega_3,\omega_4)
  &=\frac{1}{Z}
   \!\!\sum_{\substack{jlfp \cr \varepsilon_j\neq\varepsilon_f \cr \varepsilon_p\neq\varepsilon_l}}\!\! \rme^ {-\beta\varepsilon_j} A_{jl}B_{lf}C_{fp}D_{pj}
   \delta(\varepsilon_{jl}+\omega_1)\delta(\varepsilon_{lf}+\omega_2)\delta(\varepsilon_{fp}+\omega_3),
\label{SD_4t_cont}\\
 \bar{I}_{\overline{AB}\;\overline{CD}}(\omega_1,-\omega_1,\omega_3,-\omega_3)
   &= \frac{1}{Z}
   \!\!\sum_{\substack{jlfp \cr \varepsilon_j=\varepsilon_f \cr \varepsilon_p\neq\varepsilon_l}}\!\! \rme^ {-\beta\varepsilon_j} A_{jl}B_{lf}C_{fp}D_{pj}
   \delta(\varepsilon_{jl}+\omega_1)\delta(\varepsilon_{fp}+\omega_3),
\label{SD_4t_cont1}\\
 \bar{I}_{\overline{A}\;\overline{BC}\;\overline{D}}(\omega_1,-\omega_3,\omega_3,-\omega_1)
  &= \frac{1}{Z}
  \!\!\sum_{\substack{jlfp \cr \varepsilon_l=\varepsilon_p \cr \varepsilon_j\neq\varepsilon_f}}\!\! \rme^ {-\beta\varepsilon_j} A_{jl}B_{lf}C_{fp}D_{pj}
   \delta(\varepsilon_{jl}+\omega_1)\delta(\varepsilon_{fp}+\omega_3),
\label{SD_4t_cont2}\\
 \bar{\bar{I}}_{\overline{ABCD}}(\omega_1,-\omega_1,\omega_1,-\omega_1)  &= \frac{1}{Z}
   \!\!\sum_{\substack{jlfp \cr \varepsilon_f=\varepsilon_j \cr \varepsilon_l=\varepsilon_p}}\!\!
   \rme^ {-\beta\varepsilon_j} A_{jl}B_{lf}C_{fp}D_{pj}
                    \delta(\varepsilon_{jl}+\omega_1).
        \label{SD_4t_cont3}
\end{align}
For bosonic operators, additional terms with $\delta(\omega_i)$ could appear \cite{shvaika:447}. Expression \eref{SD_4t} already displays the zero-frequency anomaly --- the presence of terms with $\delta$-function factors, which results in different time dependences of the contributions in the correlation function
\begin{align}
K_{ABCD}(t_1,t_2,t_3,t_4)&=\tilde{K}_{ABCD}(t_1,t_2,t_3,t_4)
+\bar{K}_{\overline{AB}\;\overline{CD}}(t_1-t_2,t_3-t_4)
\nonumber\\
&+\bar{K}_{\overline{A}\;\overline{BC}\;\overline{D}}(t_1-t_4,t_3-t_2)
+\bar{\bar{K}}_{\overline{ABCD}}(t_1-t_2+t_3-t_4)
\label{CF_4t_tot}
\end{align}
with a different asymptotic behavior at large time values $t_1\to\infty$, $t_2\to\infty$, $t_3\to\infty$, and $t_4\to\infty$. The first term always goes to zero $\tilde K_{ABCD}(t_1,t_2,t_3,t_4)\to0$, the next two terms $\bar{K}_{\overline{AB}\;\overline{CD}}(t_1-t_2,t_3-t_4)$ and $\bar{K}_{\overline{A}\;\overline{BC}\;\overline{D}}(t_1-t_4,t_3-t_2)$ are finite for finite differences $|t_1-t_2|\ll\infty$, $|t_3-t_4|\ll\infty$ and $|t_1-t_4|\ll\infty$, $|t_3-t_2|\ll\infty$, respectively, and the last term $\bar{\bar{K}}_{\overline{ABCD}}(t_1-t_2+t_3-t_4)$ is finite for finite values of $|t_1-t_2+t_3-t_4|\ll\infty$.

Besides, the total spectral density \eref{SD_4t} satisfies the following cyclic permutation identities ($\omega_1+\omega_2+\omega_3+\omega_4=0$)
\begin{align}
    I_{ABCD}(\omega_1,\omega_2,\omega_3,\omega_4)
   & =I_{BCDA}(\omega_2,\omega_3,\omega_4,\omega_1) \rme^ {\beta\omega_1}
    =I_{CDAB}(\omega_3,\omega_4,\omega_1,\omega_2) \rme^ {\beta(\omega_1+\omega_2)}
\nonumber\\
    &=I_{DABC}(\omega_4,\omega_1,\omega_2,\omega_3) \rme^ {-\beta\omega_4},
    \label{CF4tCicle}
\end{align}
and for a given set of operators $\hat{A}$, $\hat{B}$, $\hat{C}$, and $\hat{D}$ there are $4!=24$ different correlation functions \eref{CF_4t_tot} but only $3!=6$ nonidentical spectral densities \eref{SD_4t}.

\section{Four-time Matsubara Green's function}\label{sec:4Matsu}

Now we introduce four-time Matsubara Green's function
\begin{align}
  K_{c}^{(4)}(\tau_1,\tau_2,\tau_3,\tau_4)&=\langle \mathcal{T} \hat{A}(\tau_1) \hat{B}(\tau_2) \hat{C}(\tau_3) \hat{D}(\tau_4)\rangle,
  \nonumber\\
  K_{c}^{(4)}(\tau_1,\tau_2,\tau_3,\tau_4)&=K_{c}^{(4)}(\tau_1-\tau,\tau_2-\tau,\tau_3-\tau,\tau_4-\tau).
\end{align}
Due to the imaginary time ordering $\mathcal{T}$, its Fourier transform contains $4!=24$ terms which can be collected into $3!=6$ contributions
\begin{align}
 & K_{c}^{(4)}(\rmi\omega_{n_1},\rmi\omega_{n_2},\rmi\omega_{n_3},\rmi\omega_{n_4}) = \frac{1}{\beta}\!\int\limits_{0}^{\beta}\!\rmd\tau_1
   \!\int\limits_{0}^{\beta}\!\rmd\tau_2 \!\int\limits_{0}^{\beta}\!\rmd\tau_3 \!\int\limits_{0}^{\beta}\!\rmd\tau_4
          \rme^ {(\rmi\omega_{n_1}\tau_1+\rmi\omega_{n_2}\tau_2+\rmi\omega_{n_3}\tau_3+\rmi\omega_{n_4}\tau_4)}K_{c}^{(4)}(\tau_1,\tau_2,\tau_3,\tau_4)
 \nonumber\\
 & = \mathfrak{K}_{ABCD}(\rmi\omega_{n_1},\rmi\omega_{n_2},\rmi\omega_{n_3},\rmi\omega_{n_4})
  + \mathfrak{K}_{DCBA}(\rmi\omega_{n_4},\rmi\omega_{n_3},\rmi\omega_{n_2},\rmi\omega_{n_1})
 + \mathfrak{K}_{ACDB}(\rmi\omega_{n_1},\rmi\omega_{n_3},\rmi\omega_{n_4},\rmi\omega_{n_2})
 \nonumber\\
 & + \mathfrak{K}_{BDCA}(\rmi\omega_{n_2},\rmi\omega_{n_4},\rmi\omega_{n_3},\rmi\omega_{n_1})
  + \mathfrak{K}_{ADBC}(\rmi\omega_{n_1},\rmi\omega_{n_4},\rmi\omega_{n_2},\rmi\omega_{n_3})
  + \mathfrak{K}_{CBDA}(\rmi\omega_{n_3},\rmi\omega_{n_2},\rmi\omega_{n_4},\rmi\omega_{n_1}),
            \label{K4tFT}
\end{align}
where
\begin{equation}
\mathfrak{K}_{ABCD}(\rmi\omega_{n_1},\rmi\omega_{n_2},\rmi\omega_{n_3},\rmi\omega_{n_4})
= \frac{1}{Z}\!\sum_{jlfp}\! A_{jl}B_{lf}C_{fp}D_{pj}\mathfrak{P}(j,\rmi\omega_{n_1},l,\rmi\omega_{n_2},f,\rmi\omega_{n_3},p,\rmi\omega_{n_4})
\label{K4FTcyc}
\end{equation}
collects the terms connected by cyclic permutations and $\rmi\omega_{n}=\rmi(2n+1)\pi T$ are fermionic Matsubara frequencies which satisfy the constraint
\begin{equation}\label{matsfreqconserv}
\rmi\omega_{n_1}+\rmi\omega_{n_2}+\rmi\omega_{n_3}+\rmi\omega_{n_4}=0.
\end{equation}

In equation~\eref{K4FTcyc}, cyclic permutations are included through quantity
\begin{align}
     \mathfrak{P}(j,\rmi\omega_{n_1},l,\rmi\omega_{n_2},f,\rmi\omega_{n_3},&p,\rmi\omega_{n_4})
     =\frac{1}{\beta}\Biggl[
     \rme^ {-\beta\varepsilon_j}\int\limits_{0}^{\beta}\rmd\tau_1 \int\limits_{0}^{\tau_1}\rmd\tau_2
                               \int\limits_{0}^{\tau_2}\rmd\tau_3 \int\limits_{0}^{\tau_3}\rmd\tau_4
   -\rme^ {-\beta\varepsilon_l}\int\limits_{0}^{\beta}\rmd\tau_2 \int\limits_{0}^{\tau_2}\rmd\tau_3
                               \int\limits_{0}^{\tau_3}\rmd\tau_4 \int\limits_{0}^{\tau_4}\rmd\tau_1
     \nonumber\\
   &  +\rme^ {-\beta\varepsilon_f}\int\limits_{0}^{\beta}\rmd\tau_3 \int\limits_{0}^{\tau_3}\rmd\tau_4
                               \int\limits_{0}^{\tau_4}\rmd\tau_1 \int\limits_{0}^{\tau_1}\rmd\tau_2
  -\rme^ {-\beta\varepsilon_p}\int\limits_{0}^{\beta}\rmd\tau_4 \int\limits_{0}^{\tau_4}\rmd\tau_1
                               \int\limits_{0}^{\tau_1}\rmd\tau_2 \int\limits_{0}^{\tau_2}\rmd\tau_3
     \Biggr]
     \nonumber\\
  &   \times\exp[(\varepsilon_{jl}+\rmi\omega_{n_1})\tau_1+(\varepsilon_{lf}+\rmi\omega_{n_2})\tau_2
+(\varepsilon_{fp}+\rmi\omega_{n_3})\tau_3+(\varepsilon_{pj}+\rmi\omega_{n_4})\tau_4],
     \label{P_int}
\end{align}
which satisfies an obvious relation
\begin{equation}
	\mathfrak{P}(j,\rmi\omega_{n_1},l,\rmi\omega_{n_2},f,\rmi\omega_{n_3},p,\rmi\omega_{n_4})
= -\mathfrak{P}(l,\rmi\omega_{n_2},f,\rmi\omega_{n_3},p,\rmi\omega_{n_4},j,\rmi\omega_{n_1}).
\end{equation}
In a general case, when all possible nontrivial sums of Matsubara frequencies are nonzero or when there are no eigenstates with the same energy values, function \eref{P_int} is equal to
\begin{align}
& \widetilde{\mathfrak{P}}(j,\rmi\omega_{n_1},l,\rmi\omega_{n_2},f,\rmi\omega_{n_3},p,\rmi\omega_{n_4})
    = \Delta(\rmi\omega_{n_1}{+}\rmi\omega_{n_2}{+}\rmi\omega_{n_3}{+}\rmi\omega_{n_4})
\nonumber\\
& \times\biggl[
  \frac{\rme^ {-\beta\varepsilon_j}}{(\varepsilon_{lj}-\rmi\omega_{n_1})(\varepsilon_{fj}-\rmi\omega_{n_1}-\rmi\omega_{n_2})(\varepsilon_{pj}+\rmi\omega_{n_4})}
  -\frac{\rme^ {-\beta\varepsilon_l}}{(\varepsilon_{fl}-\rmi\omega_{n_2})(\varepsilon_{pl}-\rmi\omega_{n_2}-\rmi\omega_{n_3})(\varepsilon_{jl}+\rmi\omega_{n_1})}
\nonumber\\
&  +\frac{\rme^ {-\beta\varepsilon_f}}{(\varepsilon_{if}-\rmi\omega_{n_3})(\varepsilon_{jf}-\rmi\omega_{n_3}-\rmi\omega_{n_4})(\varepsilon_{lf}+\rmi\omega_{n_2})}
   -\frac{\rme^ {-\beta\varepsilon_p}}{(\varepsilon_{jp}-\rmi\omega_{n_4})(\varepsilon_{lp}-\rmi\omega_{n_4}-\rmi\omega_{n_1})(\varepsilon_{fp}+\rmi\omega_{n_3})}
     \biggr].
     \label{P_int_gen}
\end{align}
Besides, we should consider several special cases, when we have levels with the same energy value: the case of $\varepsilon_{j}=\varepsilon_{f}\neq\varepsilon_{p},\varepsilon_{l}$ and $\rmi\omega_{n_1}+\rmi\omega_{n_2}=-\rmi\omega_{n_3}-\rmi\omega_{n_4}=0$. Now we have an additional contribution
\begin{align}
  &   \left[\lim_{\varepsilon_{f}\to\varepsilon_{j}} \lim_{\rmi\omega_{n_2}\to-\rmi\omega_{n_1}}
     - \lim_{\rmi\omega_{n_2}\to-\rmi\omega_{n_1}} \lim_{\varepsilon_{f}\to\varepsilon_{j}}\right]
     \widetilde{\mathfrak{P}}(j,\rmi\omega_{n_1},l,\rmi\omega_{n_2},f,\rmi\omega_{n_3},p,\rmi\omega_{n_4})
\nonumber\\
  &   =\Delta(\rmi\omega_{n_1}+\rmi\omega_{n_2})\Delta(\rmi\omega_{n_3}+\rmi\omega_{n_4})
     \Delta_{\varepsilon_{j},\varepsilon_{f}}
     \frac{\beta \rme^ {-\beta\varepsilon_j}}{(\varepsilon_{lj}-\rmi\omega_{n_1})(\varepsilon_{pj}-\rmi\omega_{n_3})}\,.
     \label{P_int_1}
\end{align}
Another case of $\varepsilon_{p}=\varepsilon_{l}\neq\varepsilon_{j},\varepsilon_{f}$ and $\rmi\omega_{n_1}+\rmi\omega_{n_4}=-\rmi\omega_{n_3}-\rmi\omega_{n_2}=0$ produces a different additional contribution
\begin{align}
  &   \left[\lim_{\varepsilon_{p}\to\varepsilon_{l}} \lim_{\rmi\omega_{n_4}\to-\rmi\omega_{n_1}}
     - \lim_{\rmi\omega_{n_4}\to-\rmi\omega_{n_1}} \lim_{\varepsilon_{p}\to\varepsilon_{l}}\right]
     \widetilde{\mathfrak{P}}(j,\rmi\omega_{n_1},l,\rmi\omega_{n_2},f,\rmi\omega_{n_3},p,\rmi\omega_{n_4})
\nonumber\\
  &   =-\Delta(\rmi\omega_{n_1}+\rmi\omega_{n_4})\Delta(\rmi\omega_{n_3}+\rmi\omega_{n_2})
     \Delta_{\varepsilon_{p},\varepsilon_{l}}
     \frac{\beta \rme^ {-\beta\varepsilon_l}}{(\varepsilon_{fl}-\rmi\omega_{n_2})(\varepsilon_{jl}-\rmi\omega_{n_4})}\,.
     \label{P_int_2}
\end{align}
The case of $\varepsilon_{j}=\varepsilon_{f}\neq\varepsilon_{p}=\varepsilon_{l}$ and $\rmi\omega_{n_1}=-\rmi\omega_{n_2}=\rmi\omega_{n_3}=-\rmi\omega_{n_4}$ does not introduce any additional contributions but it should be considered separately to avoid double counting.
Here,
 \begin{equation}
     \Delta_{\varepsilon_j,\varepsilon_f}=\left\{\begin{array}{ll}
    1, \quad \varepsilon_j=\varepsilon_f\\
    0, \quad \varepsilon_j\neq\varepsilon_f
    \end{array}
    \right.; \qquad  \bar\Delta_{\varepsilon_j,\varepsilon_f}=1-\Delta_{\varepsilon_j,\varepsilon_f}\,.
 \end{equation}
Special consideration of such terms is required because in many cases, e.g., in numerical calculations, it is very difficult to tune up independently the energies of each many-body state $\varepsilon_j$ in order to apply the tricks like \eref{P_int_1} and \eref{P_int_2}, and they should be incorporated in the theory explicitly. On the other hand, they correspond to the cases when the consequent action of two fermionic operators returns the many-body system back to the initial state or to the state with the same energy (true or accidental degeneracy) and represent the elastic scattering collisions. Such processes determine the difference between the isothermal (e.g., static) and isolated (Kubo) susceptibilities \cite{wilcox:624}.

Finally, we get
\begin{align}
   &  \mathfrak{P}(j,\rmi\omega_{n_1},l,\rmi\omega_{n_2},f,\rmi\omega_{n_3},p,\rmi\omega_{n_4})
  = \widetilde{\mathfrak{P}}(j,\rmi\omega_{n_1},l,\rmi\omega_{n_2},f,\rmi\omega_{n_3},p,\rmi\omega_{n_4})
\nonumber\\
  &   +\Delta(\rmi\omega_{n_1}+\rmi\omega_{n_2})\Delta(\rmi\omega_{n_3}+\rmi\omega_{n_4})
     \Delta_{\varepsilon_{j},\varepsilon_{f}}
   \frac{\beta \rme^ {-\beta\varepsilon_j}}{(\varepsilon_{lj}-\rmi\omega_{n_1})(\varepsilon_{pj}-\rmi\omega_{n_3})}
\nonumber\\
  &   -\Delta(\rmi\omega_{n_1}+\rmi\omega_{n_4})\Delta(\rmi\omega_{n_3}+\rmi\omega_{n_2})
     \Delta_{\varepsilon_{p},\varepsilon_{l}}
   \frac{\beta \rme^ {-\beta\varepsilon_l}}{(\varepsilon_{fl}-\rmi\omega_{n_2})(\varepsilon_{jl}-\rmi\omega_{n_4})}
     \label{P_int_tot}
\end{align}
or
\begin{align}
 &    \mathfrak{P}(j,\rmi\omega_{n_1},l,\rmi\omega_{n_2},f,\rmi\omega_{n_3},p,\rmi\omega_{n_4})
     =\Delta(\rmi\omega_{n_1}+\rmi\omega_{n_2}+\rmi\omega_{n_3}+\rmi\omega_{n_4})
     \bar\Delta_{\varepsilon_{j},\varepsilon_{f}}\bar\Delta_{\varepsilon_{l},\varepsilon_{p}}
\nonumber\\
 &    \times\biggl[
     \frac{\rme^ {-\beta\varepsilon_j}}{(\varepsilon_{lj}-\rmi\omega_{n_1})(\varepsilon_{fj}-\rmi\omega_{n_1}-\rmi\omega_{n_2})(\varepsilon_{pj}+\rmi\omega_{n_4})}
     - \frac{\rme^ {-\beta\varepsilon_l}}{(\varepsilon_{fl}-\rmi\omega_{n_2})(\varepsilon_{pl}-\rmi\omega_{n_2}-\rmi\omega_{n_3})(\varepsilon_{jl}+\rmi\omega_{n_1})}
\nonumber\\
  &   + \frac{\rme^ {-\beta\varepsilon_f}}{(\varepsilon_{pf}-\rmi\omega_{n_3})(\varepsilon_{jf}-\rmi\omega_{n_3}-\rmi\omega_{n_4})(\varepsilon_{lf}+\rmi\omega_{n_2})}
     - \frac{\rme^ {-\beta\varepsilon_p}}{(\varepsilon_{jp}-\rmi\omega_{n_4})(\varepsilon_{lp}-\rmi\omega_{n_4}-\rmi\omega_{n_1})(\varepsilon_{fp}+\rmi\omega_{n_3})}
     \biggr]
\nonumber\\
  &   -\Delta(\rmi\omega_{n_1}+\rmi\omega_{n_2}+\rmi\omega_{n_3}+\rmi\omega_{n_4})
     \Delta_{\varepsilon_{j},\varepsilon_{f}}\bar\Delta_{\varepsilon_{l},\varepsilon_{p}}
   \biggl\{
     \frac{\rme^ {-\beta\varepsilon_j}}{(\varepsilon_{lj}+\rmi\omega_{n_2})(\varepsilon_{pj}+\rmi\omega_{n_4})}
     \biggl[\frac{1}{\varepsilon_{lj}-\rmi\omega_{n_1}}+\frac{1}{\varepsilon_{pj}-\rmi\omega_{n_3}} \biggr]
\nonumber\\
  &   +\frac{1}{\varepsilon_{pl}-\rmi\omega_{n_2}-\rmi\omega_{n_3}} \biggl[ \frac{\rme^ {-\beta\varepsilon_l}}{(\varepsilon_{jl}-\rmi\omega_{n_2})(\varepsilon_{jl}+\rmi\omega_{n_1})}
     - \frac{\rme^ {-\beta\varepsilon_p}}{(\varepsilon_{jp}-\rmi\omega_{n_4})(\varepsilon_{jp}+\rmi\omega_{n_3})}\biggr]
     \biggr\}
\nonumber\\
  &   +\Delta(\rmi\omega_{n_1}+\rmi\omega_{n_2}+\rmi\omega_{n_3}+\rmi\omega_{n_4})
     \bar\Delta_{\varepsilon_{j},\varepsilon_{f}}\Delta_{\varepsilon_{l},\varepsilon_{p}}
\nonumber\\
  &   \times\biggl\{
     \frac{1}{\varepsilon_{fj}-\rmi\omega_{n_1}-\rmi\omega_{n_2}} \biggl[ \frac{\rme^ {-\beta\varepsilon_j}}{(\varepsilon_{lj}-\rmi\omega_{n_1})(\varepsilon_{lj}+\rmi\omega_{n_4})}
          - \frac{\rme^ {-\beta\varepsilon_f}}{(\varepsilon_{lf}-\rmi\omega_{n_3})(\varepsilon_{lf}+\rmi\omega_{n_2})}\biggr]
\nonumber\\
  &   +\frac{\rme^ {-\beta\varepsilon_l}}{(\varepsilon_{jl}+\rmi\omega_{n_1})(\varepsilon_{fl}+\rmi\omega_{n_3})}
          \biggl[\frac{1}{\varepsilon_{fl}-\rmi\omega_{n_2}}+\frac{1}{\varepsilon_{jl}-\rmi\omega_{n_4}} \biggr]
          \biggr\}
\nonumber\\
   &  +\Delta(\rmi\omega_{n_1}+\rmi\omega_{n_2}+\rmi\omega_{n_3}+\rmi\omega_{n_4})
     \Delta_{\varepsilon_{j},\varepsilon_{f}}\Delta_{\varepsilon_{l},\varepsilon_{p}}
     \frac{\rme^ {-\beta\varepsilon_j}+\rme^ {-\beta\varepsilon_l}}{(\varepsilon_{lj}-\rmi\omega_{n_1})(\varepsilon_{lj}-\rmi\omega_{n_3})}
  \biggl[\frac{1}{\varepsilon_{jl}-\rmi\omega_{n_2}}+\frac{1}{\varepsilon_{jl}-\rmi\omega_{n_4}} \biggr]
\nonumber\\
  &   +\Delta(\rmi\omega_{n_1}+\rmi\omega_{n_2})\Delta(\rmi\omega_{n_3}+\rmi\omega_{n_4})
     \Delta_{\varepsilon_{j},\varepsilon_{f}}\bar\Delta_{\varepsilon_{p},\varepsilon_{l}}
     \frac{\beta \rme^ {-\beta\varepsilon_j}}{(\varepsilon_{lj}-\rmi\omega_{n_1})(\varepsilon_{pj}-\rmi\omega_{n_3})}
\nonumber\\
   &  -\Delta(\rmi\omega_{n_1}+\rmi\omega_{n_4})\Delta(\rmi\omega_{n_3}+\rmi\omega_{n_2})
     \bar\Delta_{\varepsilon_{j},\varepsilon_{f}}\Delta_{\varepsilon_{p},\varepsilon_{l}}
     \frac{\beta \rme^ {-\beta\varepsilon_l}}{(\varepsilon_{fl}-\rmi\omega_{n_2})(\varepsilon_{jl}-\rmi\omega_{n_4})}.
\nonumber\\
  &   +
     \frac{\beta\Delta_{\varepsilon_{j},\varepsilon_{f}}\Delta_{\varepsilon_{p},\varepsilon_{l}}}{(\varepsilon_{lj}-\rmi\omega_{n_1})(\varepsilon_{lj}-\rmi\omega_{n_3})}
     \bigl[\Delta(\rmi\omega_{n_1}+\rmi\omega_{n_2})\Delta(\rmi\omega_{n_3}+\rmi\omega_{n_4}) \rme^ {-\beta\varepsilon_j}
  -\Delta(\rmi\omega_{n_1}+\rmi\omega_{n_4})\Delta(\rmi\omega_{n_3}+\rmi\omega_{n_2}) \rme^ {-\beta\varepsilon_l}\bigr].
     \label{P_int_tot_detail}
\end{align}

Now we can introduce spectral representations for four-time fermionic Matsubara functions. For the first term in \eref{K4tFT} we get ($\omega_4=-\omega_1-\omega_2-\omega_3$)
\begin{align}
&\mathfrak{K}_{ABCD}(\rmi\omega_{n_1},\rmi\omega_{n_2},\rmi\omega_{n_3},\rmi\omega_{n_4})
= \widetilde{\mathfrak{K}}_{ABCD}(\rmi\omega_{n_1},\rmi\omega_{n_2},\rmi\omega_{n_3},\rmi\omega_{n_4})
\nonumber\\
&+\beta\Delta(\rmi\omega_{n_1}+\rmi\omega_{n_2})\Delta(\rmi\omega_{n_3}+\rmi\omega_{n_4})
\!\!\int\limits_{-\infty}^{+\infty}\!\! \rmd\omega_1  \!\!\int\limits_{-\infty}^{+\infty}\!\! \rmd\omega_3
\frac{\bar{I}_{\overline{AB}\;\overline{CD}}(\omega_1,-\omega_1,\omega_3,-\omega_3)}{(\omega_1-\rmi\omega_{n_1})(\omega_3-\rmi\omega_{n_3})}
\nonumber\\
&-\beta\Delta(\rmi\omega_{n_1}+\rmi\omega_{n_4})\Delta(\rmi\omega_{n_3}+\rmi\omega_{n_2})
\!\!\int\limits_{-\infty}^{+\infty}\!\! \rmd\omega_1  \!\!\int\limits_{-\infty}^{+\infty}\!\! \rmd\omega_3
\frac{\bar{I}_{\overline{A}\;\overline{BC}\;\overline{D}}(\omega_1,-\omega_3,\omega_3,-\omega_1)}{(\omega_1-\rmi\omega_{n_1})(\omega_3-\rmi\omega_{n_3})}
\rme^ {-\beta\omega_1}
\nonumber\\
&+\beta\Delta(\rmi\omega_{n_1}+\rmi\omega_{n_2})\Delta(\rmi\omega_{n_3}+\rmi\omega_{n_4})
\!\!\int\limits_{-\infty}^{+\infty}\!\! \rmd\omega_1
\frac{\bar{\bar{I}}_{\overline{ABCD}}(\omega_1,-\omega_1,\omega_1,-\omega_1)}{(\omega_1-\rmi\omega_{n_1})(\omega_1-\rmi\omega_{n_3})}
\nonumber\\
&-\beta\Delta(\rmi\omega_{n_1}+\rmi\omega_{n_4})\Delta(\rmi\omega_{n_3}+\rmi\omega_{n_2})
\!\!\int\limits_{-\infty}^{+\infty}\!\! \rmd\omega_1
\frac{\bar{\bar{I}}_{\overline{ABCD}}(\omega_1,-\omega_1,\omega_1,-\omega_1)}{(\omega_1-\rmi\omega_{n_1})(\omega_1-\rmi\omega_{n_3})}
\rme^ {-\beta\omega_1},
\label{eq:4timelehman}
\end{align}
where
\begin{align}
&\widetilde{\mathfrak{K}}_{ABCD}(\rmi\omega_{n_1},\rmi\omega_{n_2},\rmi\omega_{n_3},\rmi\omega_{n_4})=\Delta(\rmi\omega_{n_1}+\rmi\omega_{n_2}+\rmi\omega_{n_3}+\rmi\omega_{n_4})
\Biggl(\int\limits_{-\infty}^{+\infty}\!\! \rmd\omega_1 \!\!\int\limits_{-\infty}^{+\infty}\!\! \rmd\omega_2 \!\!\int\limits_{-\infty}^{+\infty}\!\! \rmd\omega_3 \tilde{I}_{ABCD}(\omega_1,\omega_2,\omega_3,\omega_4)
\nonumber\\
&\times\Biggl[
\frac{1}{(\omega_1-\rmi\omega_{n_1})(\omega_1+\omega_2-\rmi\omega_{n_1}-\rmi\omega_{n_2})(-\omega_4+\rmi\omega_{n_4})}
-\frac{\rme^ {-\beta\omega_1}}{(\omega_2-\rmi\omega_{n_2})(\omega_2+\omega_3-\rmi\omega_{n_2}-\rmi\omega_{n_3})(-\omega_1+\rmi\omega_{n_1})}
\nonumber\\
&+\frac{\rme^ {-\beta(\omega_1+\omega_2)}}{(\omega_3-\rmi\omega_{n_3})(\omega_3+\omega_4-\rmi\omega_{n_3}-\rmi\omega_{n_4})(-\omega_2+\rmi\omega_{n_2})}
-\frac{\rme^ {\beta\omega_4}}{(\omega_4-\rmi\omega_{n_4})(\omega_4+\omega_1-\rmi\omega_{n_4}-\rmi\omega_{n_1})(-\omega_3+\rmi\omega_{n_3})}
\Biggr]
\nonumber\\
&-\!\!\int\limits_{-\infty}^{+\infty}\!\! \rmd\omega_1  \!\!\int\limits_{-\infty}^{+\infty}\!\! \rmd\omega_3
\bar{I}_{\overline{AB}\;\overline{CD}}(\omega_1,-\omega_1,\omega_3,-\omega_3)
\Biggl\{
\frac{1}{(\omega_1+\rmi\omega_{n_2})(\omega_3+\rmi\omega_{n_4})} \Biggl[\frac{1}{\omega_1-\rmi\omega_{n_1}}+\frac{1}{\omega_3-\rmi\omega_{n_3}}\Biggr]
\nonumber\\
&+\frac{1}{\omega_3-\omega_1-\rmi\omega_{n_2}-\rmi\omega_{n_3}} \Biggl[\frac{\rme^ {-\beta\omega_1}}{(\omega_1+\rmi\omega_{n_2})(\omega_1-\rmi\omega_{n_1})} - \frac{\rme^ {-\beta\omega_3}}{(\omega_3+\rmi\omega_{n_4})(\omega_3-\rmi\omega_{n_3})} \Biggr]
\Biggr\}
\nonumber\\
&+\!\!\int\limits_{-\infty}^{+\infty}\!\! \rmd\omega_1  \!\!\int\limits_{-\infty}^{+\infty}\!\! \rmd\omega_3
\bar{I}_{\overline{A}\;\overline{BC}\;\overline{D}}(\omega_1,-\omega_3,\omega_3,-\omega_1)
\Biggl\{
\frac{\rme^ {-\beta\omega_1}}{(\omega_1-\rmi\omega_{n_1})(\omega_3-\rmi\omega_{n_3})} \Biggl[\frac{1}{-\omega_3-\rmi\omega_{n_2}}+\frac{1}{-\omega_1-\rmi\omega_{n_4}}\Biggr]
\nonumber\\
&+\frac{1}{\omega_1-\omega_3-\rmi\omega_{n_1}-\rmi\omega_{n_2}} \Biggl[\frac{1}{(\omega_1+\rmi\omega_{n_4})(\omega_1-\rmi\omega_{n_1})} - \frac{\rme^ {-\beta(\omega_1-\omega_3)}}{(\omega_3+\rmi\omega_{n_2})(\omega_3-\rmi\omega_{n_3})} \Biggr]
\Biggr\}
\nonumber\\
&-\!\!\int\limits_{-\infty}^{+\infty}\!\! \rmd\omega_1
\bar{\bar{I}}_{\overline{ABCD}}(\omega_1,-\omega_1,\omega_1,-\omega_1)
\frac{1+\rme^ {-\beta\omega_1}}{(\omega_1-\rmi\omega_{n_1})(\omega_1-\rmi\omega_{n_3})}
\Biggl[\frac{1}{\omega_1+\rmi\omega_{n_2}}+\frac{1}{\omega_1+\rmi\omega_{n_4}}\Biggr]\Biggr).
\label{eq:4timeanomal}
\end{align}
Similar expressions can be written for the other five contributions in \eref{K4tFT}. One can see from \eref{eq:4timelehman} and \eref{eq:4timeanomal} that spectral densities \eref{SD_4t_cont} contribute only in the normal components \eref{eq:4timeanomal}, whereas the spectral densities \eref{SD_4t_cont1}--\eref{SD_4t_cont3} contribute both in the normal and in anomalous components.

\subsection{Zero-frequency anomaly and cumulants}

It follows from equation~\eref{eq:4timelehman} that there are two types of contributions: the normal one $\widetilde{\mathfrak{K}}_{ABCD}(\rmi\omega_{n_1},\rmi\omega_{n_2},\rmi\omega_{n_3},\rmi\omega_{n_4})$ with all frequencies being different and the anomalous one with additional constraints on the frequencies. Based on this, one can rewrite the four-time Green's function \eref{K4tFT} in the form
\begin{align}
	 K_{c}^{(4)}(\rmi\omega_{n_1},\rmi\omega_{n_2},\rmi\omega_{n_3},\rmi\omega_{n_4}) &= \widetilde{K}_{c}^{(4)}(\rmi\omega_{n_1},\rmi\omega_{n_2},\rmi\omega_{n_3},\rmi\omega_{n_4})
	\nonumber\\
	& + \beta\Delta(\rmi\omega_{n_1}+\rmi\omega_{n_2})\Delta(\rmi\omega_{n_3}+\rmi\omega_{n_4}) \bar{K}_{c}^{AB,CD}(\rmi\omega_{n_1},\rmi\omega_{n_3})
	\nonumber\\
	& + \beta\Delta(\rmi\omega_{n_1}+\rmi\omega_{n_3})\Delta(\rmi\omega_{n_4}+\rmi\omega_{n_2}) \bar{K}_{c}^{AC,DB}(\rmi\omega_{n_1},\rmi\omega_{n_4})
	\nonumber\\
	& + \beta\Delta(\rmi\omega_{n_1}+\rmi\omega_{n_4})\Delta(\rmi\omega_{n_2}+\rmi\omega_{n_3}) \bar{K}_{c}^{AD,BC}(\rmi\omega_{n_1},\rmi\omega_{n_2}),
\label{K4tFTrecol}
\end{align}
where
\begin{align}
	\widetilde{K}_{c}^{(4)}(\rmi\omega_{n_1},\rmi\omega_{n_2},\rmi\omega_{n_3},\rmi\omega_{n_4})
	& = \widetilde{\mathfrak{K}}_{ABCD}(\rmi\omega_{n_1},\rmi\omega_{n_2},\rmi\omega_{n_3},\rmi\omega_{n_4})
	 + \widetilde{\mathfrak{K}}_{DCBA}(\rmi\omega_{n_4},\rmi\omega_{n_3},\rmi\omega_{n_2},\rmi\omega_{n_1})
	\nonumber\\
	& + \widetilde{\mathfrak{K}}_{ACDB}(\rmi\omega_{n_1},\rmi\omega_{n_3},\rmi\omega_{n_4},\rmi\omega_{n_2})
	 + \widetilde{\mathfrak{K}}_{BDCA}(\rmi\omega_{n_2},\rmi\omega_{n_4},\rmi\omega_{n_3},\rmi\omega_{n_1})
	\nonumber\\
	& + \widetilde{\mathfrak{K}}_{ADBC}(\rmi\omega_{n_1},\rmi\omega_{n_4},\rmi\omega_{n_2},\rmi\omega_{n_3})
	 + \widetilde{\mathfrak{K}}_{CBDA}(\rmi\omega_{n_3},\rmi\omega_{n_2},\rmi\omega_{n_4},\rmi\omega_{n_1})
	\label{K4tFTreg}
\end{align}
collects all normal contributions. The anomalous contribution has the form:
\begin{equation}\label{eq:K4anom}
	\bar{K}_{c}^{AB,CD}(\rmi\omega_{n_1},\rmi\omega_{n_3}) = \frac{1}{Z} \sum_{\substack{jf \cr \varepsilon_j=\varepsilon_f}} \rme^ {-\beta\varepsilon_j} g_{jf}^{AB}(\rmi\omega_{n_1}) g_{fj}^{CD}(\rmi\omega_{n_3}),
\end{equation}
where the quantities
\begin{equation}
	g_{jf}^{AB}(\rmi\omega_{n_1}) = \sum_l \left[\frac{A_{jl}B_{lf}}{\rmi\omega_1-\varepsilon_{lj}}+\frac{B_{jl}A_{lf}}{\rmi\omega_1+\varepsilon_{lj}}\right]
\end{equation}
could be considered as unaveraged matrix elements of the two-time Green's function \eref{eq:K2erg}
\begin{equation}
\widetilde{G}_{AB}(\rmi\omega_n) = \frac{1}{Z} \sum_{jf} \rme^ {-\beta\varepsilon_j}
\left[\frac{A_{jf} B_{fj}}{\rmi\omega_n+\varepsilon_{jf}} \mp \frac{B_{jf} A_{fj}}{\rmi\omega_n-\varepsilon_{jf}}\right]
= \sum_{j} b_j^{(1)} g_{jj}^{AB}(\rmi\omega_{n}).
\label{eq:G_AB}
\end{equation}
Here,
\begin{equation}
b_j^{(1)} = \frac{\partial }{\partial(-\beta\varepsilon_j)} \ln Z = \frac{1}{Z} \rme^ {-\beta\varepsilon_j}
\end{equation}
could be considered as a first cumulant (Ursell function) \cite{ursell:685,fisher:199,kubo:1100} of the Boltzmann distribution.

Anomalous term can be represented as a sum of two contributions
\begin{equation}
\bar{K}_{c}^{AB,CD}(\rmi\omega_{n_1},\rmi\omega_{n_3}) = \widetilde{G}_{AB}(\rmi\omega_{n_1}) \widetilde{G}_{CD}(\rmi\omega_{n_3})
+ \bar{K}_{c,{\rm irr}}^{AB,CD}(\rmi\omega_{n_1},\rmi\omega_{n_3}),
\end{equation}
where the reducible part [product of two two-time Green's functions \eref{eq:K2erg}] is separated and the irreducible one is equal to
\begin{equation}
	\bar{K}_{c,{\rm irr}}^{AB,CD}(\rmi\omega_{n_1},\rmi\omega_{n_3}) = \sum_{\substack{jfj'f' \cr \varepsilon_j=\varepsilon_f}} \left[ b_j^{(1)} \delta_{jj'} \delta_{ff'} - b_j^{(1)} b_{j'}^{(1)} \delta_{jf} \delta_{j'f'}\right]
	 g_{jf}^{AB}(\rmi\omega_{n_1}) g_{f'j'}^{CD}(\rmi\omega_{n_3}).
	\label{eq:K4irr}
\end{equation}
Let us consider the case of non-degenerate states, when there are no different states with the same energy value. In this case, expression \eref{eq:K4irr} takes up a straighter form
\begin{equation}
	\bar{K}_{c,{\rm irr}}^{AB,CD}(\rmi\omega_{n_1},\rmi\omega_{n_3}) = \sum_{jf} b_{jf}^{(2)} g_{jj}^{AB}(\rmi\omega_{n_1}) g_{ff}^{CD}(\rmi\omega_{n_3}),
\end{equation}
where
\begin{equation}
b_{jf}^{(2)} = \frac{\partial }{\partial(-\beta\varepsilon_j)} \frac{\partial }{\partial(-\beta\varepsilon_f)} \ln Z
= \frac{\partial b_{j}^{(1)}}{\partial(-\beta\varepsilon_f)} = \frac{\partial b_{f}^{(1)}}{\partial(-\beta\varepsilon_j)}
= b_{j}^{(1)} \delta_{jf} - b_{j}^{(1)} b_{f}^{(1)}
\end{equation}
is the second cumulant of the Boltzmann distribution. Based on this, one can consider an expression in brackets in \eref{eq:K4irr} as a generalization of cumulant expansions for the case when degenerate states are present in the many-body system.

For noninteracting fermions, the irreducible four-time Green's function \eref{eq:K4irr} is equal to zero $\bar{K}_{c,{\rm irr}}^{AB,CD}(\rmi\omega_{n_1},\rmi\omega_{n_3})=0$. For interacting fermions, there are not so many exact solutions for multi-time response functions. One solution is known for the impurity problem \cite{brandt:365} in the dynamical mean field theory \cite{metzner:324,georges:13} for the Falicov-Kimball model \cite{falicov:997,freericks:1333} with a generalized partition function
\begin{equation}
\mathfrak{Z} = \Tr \exp \left(-\beta H_{{\rm imp}}\right)
 \mathcal{T}\exp \left[-\int_{0}^{\beta} \rmd\tau \int_{0}^{\beta} \rmd\tau' \lambda(\tau-\tau') d^{\dagger}(\tau) d(\tau')\right],
\end{equation}
where the impurity Hamiltonian
\begin{equation}
H_{{\rm imp}} = U d^{\dagger} d\, f^{\dagger} f + E_1 f^{\dagger} f + E_0 f f^{\dagger} - \mu \left(d^{\dagger} d + f^{\dagger} f\right)
\end{equation}
describes the local Coulomb interaction of value $U$ between the itinerant $d$ and localized $f$-electrons and $\lambda(\tau-\tau')$ is a generalized field. Here, $E_1$ and $E_0$ are energies of the occupied and unoccupied $f$-state ($E_0\to0$), respectively, and $\mu$ is chemical potential. An occupation of the $f$-state $f^{\dagger} f$ is a conserved quantity that allows one to find the two-time Green's function \eref{eq:G_AB} in the form
\begin{equation}
G_{dd^{\dagger}}(\rmi\omega_n) = w_0 g_0(\rmi\omega_n) + w_1 g_1(\rmi\omega_n),
\end{equation}
where
\begin{align}
	g_0(\rmi\omega_n) &= \frac{1}{\rmi\omega_n + \mu - \lambda(\rmi\omega_n)}\,,
	\nonumber\\
	g_1(\rmi\omega_n) &= \frac{1}{\rmi\omega_n + \mu -U - \lambda(\rmi\omega_n)}
\end{align}
are Green's functions for the sites unoccupied and occupied by $f$-electron, respectively, and
\begin{align}
w_0=b_0^{(1)} &= \frac{\partial }{\partial(-\beta E_0)} \ln \mathfrak{Z} ,
\nonumber\\
w_1=b_1^{(1)} &= \frac{\partial }{\partial(-\beta E_1)} \ln \mathfrak{Z}
\end{align}
are the corresponding probabilities of finding such sites ($w_0+w_1=1$).

The normal contributions in the four-time Green's function \eref{K4tFTreg} vanish in this case
\begin{equation}
\widetilde{K}_{c}^{(4)}(\rmi\omega_{n_1},\rmi\omega_{n_2},\rmi\omega_{n_3},\rmi\omega_{n_4}) =0,
\end{equation}
whereas anomalous contributions \eref{eq:K4anom} are equal to
\begin{align}
\bar{K}_{c}^{dd^{\dagger},dd^{\dagger}}(\rmi\omega_{n_1},\rmi\omega_{n_3}) &= w_0 g_0(\rmi\omega_{n_1}) g_0(\rmi\omega_{n_3})
+ w_1 g_1(\rmi\omega_{n_1}) g_1(\rmi\omega_{n_3}),
\nonumber\\
\bar{K}_{c}^{dd,d^{\dagger}d^{\dagger}}(\rmi\omega_{n_1},\rmi\omega_{n_3}) &=0,
\end{align}
and for the irreducible part \eref{eq:K4irr} we get
\begin{align}
	\bar{K}_{c,{\rm irr}}^{dd^{\dagger},dd^{\dagger}}(\rmi\omega_{n_1},\rmi\omega_{n_3})
	&= \sum_{j,f=0,1} b_{jf}^{(2)} g_j(\rmi\omega_{n_1}) g_f(\rmi\omega_{n_3})
	\nonumber\\
	&=
	w_0 w_1 \left[g_0(\rmi\omega_{n_1})-g_1(\rmi\omega_{n_1})\right] \left[g_0(\rmi\omega_{n_3})-g_1(\rmi\omega_{n_3})\right]
	\nonumber \\
	&=U^2 w_0 w_1 g_0(\rmi\omega_{n_1}) g_1(\rmi\omega_{n_1}) g_0(\rmi\omega_{n_3}) g_1(\rmi\omega_{n_3}),
	\label{eq:K4irrFKM}
\end{align}
where
\begin{align}
b_{11}^{(2)} &= b_{00}^{(2)} = \frac{\partial b_1^{(1)}}{\partial(-\beta E_1)} = \frac{\partial b_0^{(1)}}{\partial(-\beta E_0)} = w_0 w_1,
\nonumber\\
b_{10}^{(2)} &= b_{01}^{(2)} = \frac{\partial b_1^{(1)}}{\partial(-\beta E_0)} = \frac{\partial b_0^{(1)}}{\partial(-\beta E_1)} = - w_0 w_1
\end{align}
are second cumulants for the $f$-state occupation probabilities. Unfortunately, the complete four-time Green's function for the lattice Falicov-Kimball model is unknown; the solutions of the Bethe-Salpeter equation in an electron-hole channel have been obtained \cite{shvaika:177,freericks:10022,shvaika:349} but these solutions break the permutation symmetry of the general expression \eref{K4tFTrecol} and more complicated approximations that include the scattering effects in different channels on equal footing, e.g., parquet one \cite{bickers:8044,janis:11345}, should be used.

In the many-body theory, cumulant contributions appear in a natural way in the strong coupling approaches \cite{hubbard:82,slobodyan:616,grewe:4420,metzner:8549,izyumov:15697,brinckmann:187} and in some cases they are the only contributions that enter the expressions for a dynamical response, e.g., dynamical charge susceptibilities \cite{shvaika:177,freericks:10022,shvaika:349} and cross-sections of the inelastic light (Raman) and x-ray scattering \cite{shvaika:137402,shvaika:045120,pakhira:125103} for the Falicov-Kimball model.

\subsection{Analytic continuation and reverse engineering problem}

Next, we can perform an analytic continuation from the Matsubara frequencies to the real one and for each term in \eref{eq:4timelehman} we will get different sets of the branch cuts as for single frequencies
\begin{equation}\label{eq:sing_cont}
  \rmi\omega_{n_{\alpha}} \rightarrow \omega_{\alpha}\pm \rmi0^{+}, \quad \alpha=1,2,3,4,
\end{equation}
as for sums of two frequencies
\begin{equation}\label{eq:pair_cont}
 \rmi\omega_{n_{\alpha}} + \rmi\omega_{n_{\gamma}} \rightarrow \omega_{\alpha} + \omega_{\gamma}\pm \rmi0^{+}.
\end{equation}
Differences in the analytic properties of each term in \eref{eq:4timelehman} and, as a result, in \eref{K4FTcyc} allow one to solve the reverse engineering problem: extracting all spectral densities from the single Green's function. To do this, one should consequently extract nonanalyticities at all brunch cuts of type \eref{eq:sing_cont} and \eref{eq:pair_cont} but in different order, which will produce a set of equations for the unknown spectral densities. For the above considered impurity problem for the Falicov-Kimball model, one can get from equation~\eref{eq:K4irrFKM} expressions for the spectral densities \eref{SD_4t_cont1} and \eref{SD_4t_cont2}, e.g.,
\begin{align}
	\bar{I}_{\overline{dd^{\dagger}}\;\overline{dd^{\dagger}}}(\omega_1,-\omega_1,\omega_3,-\omega_3)
	&= \frac{1}{\pi^2}\sum_{j,f=0,1} b_{jf}^{(2)} \Im g_j(\rmi\omega_{n_1}) \Im g_f(\rmi\omega_{n_3})
	\nonumber\\
	&=
	\frac{1}{\pi^2} w_0 w_1 \Im\left[ g_0(\rmi\omega_{n_1})- g_1(\rmi\omega_{n_1})\right] \Im\left[ g_0(\rmi\omega_{n_3})- g_1(\rmi\omega_{n_3})\right]
\end{align}
and all other spectral densities are equal to zero.
In general, the procedure is very cumbersome and will be not presented here, see for the details~\cite{shvaika:447}.

\section{High frequency asymptotics}\label{sec:highfrec}

In many applications, e.g., for the correctness checking of analytic approximations or for the memory consumption limitations for storing the high frequency tails in numerical calculations, it is useful to have the high frequency asymptotics of the four-time fermionic Matsubara Green's functions. It is obvious that for different directions in the three-frequency space defined by a constraint \eref{matsfreqconserv} one can observe a different asymptotic behavior and we shall present the results for some cases herein below.

\subsection{$|\rmi\omega_{n}|\sim\Omega$, $|\rmi\omega_{n}+\rmi\omega_{m}|\sim\Omega$, $\Omega\gg E$}

First of all we consider the most general case when each Matsubara's frequency $|\rmi\omega_{n}|\sim\Omega$ as well as each nontrivial sum of Matsubara frequencies $|\rmi\omega_{n}+\rmi\omega_{m}|\sim\Omega$ with taking into account the constraint \eref{matsfreqconserv} are much larger in modulus than the possible many-body state energy differences $\Omega\gg E=\max|\varepsilon_j-\varepsilon_l|$. The first $1/\Omega^3$ order terms in the high frequency expansion of \eref{K4tFT} using \eref{P_int_tot} are equal to
\begin{align}
& K_{c}^{(4)}(\rmi\omega_{n_1},\rmi\omega_{n_2},\rmi\omega_{n_3},\rmi\omega_{n_4}) \longrightarrow
\nonumber\\
&\frac{\langle\{[\{A,D\},B],C\}\rangle}{\rmi\omega_{n_3} \rmi\omega_{n_4} (\rmi\omega_{n_2}+\rmi\omega_{n_3})}
+ \frac{\langle\{[A,\{B,D\}],C\}\rangle}{\rmi\omega_{n_3} \rmi\omega_{n_4} (\rmi\omega_{n_1}+\rmi\omega_{n_3})} + \frac{\langle\{[C,\{A,D\}],B\}\rangle}{\rmi\omega_{n_2} \rmi\omega_{n_4} (\rmi\omega_{n_2}+\rmi\omega_{n_3})}
\nonumber\\
& + \frac{\langle\{A,[\{B,D\},C]\}\rangle}{\rmi\omega_{n_1} \rmi\omega_{n_4} (\rmi\omega_{n_1}+\rmi\omega_{n_3})}
+ \frac{\langle\{B,[\{C,D\},A]\}\rangle}{\rmi\omega_{n_2} \rmi\omega_{n_4} (\rmi\omega_{n_1}+\rmi\omega_{n_2})}
+ \frac{\langle\{A,[B,\{C,D\}]\}\rangle}{\rmi\omega_{n_1} \rmi\omega_{n_4} (\rmi\omega_{n_1}+\rmi\omega_{n_2})}\,,
\label{commut0}
\end{align}
where we have introduced anticommutators $\{X_1,X_2\}=X_1X_2+X_2X_1$ and commutators $[X,Y]=XY-YX$ of operators,
and in the case of the ordinary fermionic creation and annihilation operators it is equal to zero (but this is not correct for the Hubbard operators).

The next $1/\Omega^4$ order contributions are equal to
\begin{align}
&\mathfrak{K}_{ABCD}(\rmi\omega_{n_1},\rmi\omega_{n_2},\rmi\omega_{n_3},\rmi\omega_{n_4}) \longrightarrow
\frac{1}{\rmi\omega_{n_{1}}(\rmi\omega_{n_{1}}+\rmi\omega_{n_{2}})\rmi\omega_{n_{4}}}
\nonumber\\
&\times\left\{ \frac{\langle[A,H]BCD\rangle}{\rmi\omega_{n_{1}}} + \frac{\langle[AB,H]CD\rangle}{\rmi\omega_{n_{1}}+\rmi\omega_{n_{2}}} + \frac{\langle ABC[D,H]\rangle}{\rmi\omega_{n_{4}}} \right\},
\end{align}
where we have used the identity
\begin{equation}
\varepsilon_{lj}\langle j|\hat A|l \rangle=\langle j|[\hat A,H]|l \rangle.
\end{equation}
The presence of different frequency denominators does not allow one to collapse the total expression \eref{K4tFT} in a compact form like \eref{commut0}.

\subsection{$|\rmi\omega_{n_1}|\sim E$, $|\rmi\omega_{n}|\sim\Omega$, $|\rmi\omega_{n}+\rmi\omega_{m}|\sim\Omega$, $\Omega\gg E$}

Next, we consider the case of the finite frequency value $|\rmi\omega_{n_1}|\sim E$. Other Matsubara frequencies $|\rmi\omega_{n}|\sim\Omega$ ($n=2,3,4$) as well as each nontrivial sum of Matsubara frequencies $|\rmi\omega_{n}+\rmi\omega_{m}|\sim\Omega$ are much larger in modulus than the possible energy differences $\Omega\gg E=\max|\varepsilon_j-\varepsilon_l|$. The first terms in the high frequency expansion of \eref{K4tFT} using \eref{P_int_tot} are equal to
\newcommand*{\onefinone}[8]{
   \frac{1}{(\rmi\omega_{n_{#3}}+\rmi\omega_{n_{#4}})\rmi\omega_{n_{#4}}}
   \left\{\mathfrak{F}_{#5,#6#7#8}(\rmi\omega_{n_{#1}}) + \frac{\mathfrak{F}_{#5,#6[#7#8,H]}(\rmi\omega_{n_{#1}})}{\rmi\omega_{n_{#3}}+\rmi\omega_{n_{#4}}} + \frac{\mathfrak{F}_{#5,#6#7[#8,H]}(\rmi\omega_{n_{#1}})}{\rmi\omega_{n_{#4}}} \right\}%
   }
\newcommand*{\onefinfour}[8]{
   \frac{1}{\rmi\omega_{n_{#1}}(\rmi\omega_{n_{#1}}+\rmi\omega_{n_{#2}})}
   \left\{\mathfrak{F}_{#5#6#7,#8}(-\rmi\omega_{n_{#4}}) + \frac{\mathfrak{F}_{[#5#6,H]#7,#8}(-\rmi\omega_{n_{#4}})}{\rmi\omega_{n_{#1}}+\rmi\omega_{n_{#2}}} + \frac{\mathfrak{F}_{[#5,H]#6#7,#8}(-\rmi\omega_{n_{#4}})}{\rmi\omega_{n_{#1}}} \right\}%
   }
\newcommand*{\onefinthree}[8]{
   \frac{\langle #5#6#7#8 \rangle}{\rmi\omega_{n_{#1}}(\rmi\omega_{n_{#1}}+\rmi\omega_{n_{#2}})\rmi\omega_{n_{#4}}}
   }
\newcommand*{\onefintwo}[8]{
   \frac{\langle #5#6#7#8 \rangle}{\rmi\omega_{n_{#1}}(\rmi\omega_{n_{#1}}+\rmi\omega_{n_{#2}})\rmi\omega_{n_{#4}}}
   }
\begin{align}
& \onefinone1234ABCD
\nonumber\\
 &-\onefinone1432ADCB
\nonumber\\
 &+\onefinone1342ACDB
\nonumber\\
 &-\onefinone1243ABDC
\nonumber\\
 &+\onefinone1423ADBC
\nonumber\\
 &-\onefinone1324ACBD
\nonumber\\
 &-\onefinfour2341BCDA
\nonumber\\
 &+\onefinfour4321DCBA
\nonumber\\
 &-\onefinfour3421CDBA
\nonumber\\
 &+\onefinfour2431BDCA
\nonumber\\
 &-\onefinfour4231DBCA
\nonumber\\
 &+\onefinfour3241CBDA
\nonumber\\
 &+\onefinthree3412CDAB - \onefintwo4123DABC + \onefintwo2143BADC
\nonumber\\
 &- \onefinthree3214CBAD +\onefinthree4213DBAC - \onefintwo2134BACD
\nonumber\\
 &+ \onefintwo3124CABD - \onefinthree4312DCAB +\onefinthree2314BCAD
\nonumber\\
 &- \onefintwo3142CADB + \onefintwo4132DACB - \onefinthree2413BDAC \,.
\label{eq:hfl_12}
\end{align}
Here, we have introduced the function
\begin{equation}
\mathfrak{F}_{A,X}(\rmi\omega_{n_{1}}) = \frac{1}{Z}\sum_{jl} \rme^ {-\beta \varepsilon_j} \frac{A_{jl} X_{lj}}{\varepsilon_{lj}-\rmi\omega_{n_1}}
 = -\int\limits_{-\infty}^{+\infty} \rmd\omega n_{+}(-\omega) \frac{\rho_{AX}(\omega)}{\rmi\omega_{n_1}-\omega} \,,
\label{eq:Fhalf1}
\end{equation}
where $\rho_{AX}(\omega)$ is fermionic density of states \eref{eq:dos} which can be obtained from the corresponding Green's functions and
\(
n_{+}(\omega)={1}/\left(\rme^ {\beta\omega}+1\right)
\)
is the Fermi distribution function.
The spectral representation for function \eref{eq:Fhalf1} differs from the one for the Matsubara and the retarded (advanced) Green's functions~\eref{eq:K2erg} by the Fermi factor $n_{+}(\omega)$ and is similar to the one for the so-called ``half'' Green's functions \cite{sarry:958}.

One can imagine that the first terms in braces in \eref{eq:hfl_12} produce contributions of the order $1/\Omega^2$ whereas the other ones are of the order $1/\Omega^3$, but it could be shown that in the case of ordinary creation and annihilation fermionic operators, the total contribution of these terms is of the order $|\rmi\omega_{n_1}|/\Omega^3$, that is of the same order as the other terms are.

The cases of $|\rmi\omega_{n_2}|\sim E$, $|\rmi\omega_{n_3}|\sim E$, or $|\rmi\omega_{n_4}|\sim E$ can be obtained from the above expression by the corresponding permutation of operators and frequencies.

Next we consider the cases of the finite values of the sums of two Matsubara frequencies.

\subsection{$|\rmi\omega_{n_1}+\rmi\omega_{n_2}|\sim E$, $|\rmi\omega_{n}|\sim\Omega$, $|\rmi\omega_{n}+\rmi\omega_{m}|\sim\Omega$, $\Omega\gg E$}

First we assume that only one sum of two frequencies is finite $|\rmi\omega_{n_1}+\rmi\omega_{n_2}|=|\rmi\omega_{n_3}+\rmi\omega_{n_4}|\sim E$, including the case of $\rmi\omega_{n_1}+\rmi\omega_{n_2}=-\rmi\omega_{n_3}-\rmi\omega_{n_4}=0$. Single Matsubara frequencies $|\rmi\omega_{n}|\sim\Omega$ as well as other sums of Matsubara frequencies $|\rmi\omega_{n}+\rmi\omega_{m}|\sim\Omega$ are much larger in modulus than the possible energy differences $\Omega\gg E$. The first $1/\Omega^2$ order terms in the high frequency expansion of \eref{K4tFT} using \eref{P_int_tot_detail} are equal to
\newcommand*{\sumfinonetwo}[8]{
   \frac{\mathfrak{B}_{#7#8,#5#6}(\rmi\omega_{n_{#3}}+\rmi\omega_{n_{#4}})}{\rmi\omega_{n_{#3}}\rmi\omega_{n_{#2}}}
   +\frac{\mathfrak{B}_{#5#6,#7#8}(\rmi\omega_{n_{#1}}+\rmi\omega_{n_{#2}})}{\rmi\omega_{n_{#1}}\rmi\omega_{n_{#4}}}
   +\beta\Delta(\rmi\omega_{n_{#1}}+\rmi\omega_{n_{#2}})
    \frac{C_{#5#6,#7#8}}{\rmi\omega_{n_{#1}}\rmi\omega_{n_{#3}}}
   }
\begin{align}
 &\sumfinonetwo1234ABCD
\nonumber\\
 &+\sumfinonetwo4321DCBA
\nonumber\\
 &-\left[\sumfinonetwo3421CDBA\right]
\nonumber\\
 &-\left[\sumfinonetwo1243ABDC\right].
\end{align}
In this expression we have introduced the bosonic ``half'' Green's function \cite{sarry:958} using
\begin{equation}
\mathfrak{B}_{Y_1,Y_2}(\rmi\omega_{\nu}) = \int\limits_{-\infty}^{+\infty} \rmd\omega n_{-}(-\omega) \frac{\rho_{Y_1,Y_2}(\omega)}{\rmi\omega_{\nu}-\omega} \,,
\end{equation}
where $\rmi\omega_{\nu}=\rmi2\nu\pi T$ are bosonic Matsubara frequencies,
\(
n_{-}(\omega)={1}/\left({\rme^ {\beta\omega}-1}\right)
\)
is the Bose distribution function, $\rho_{Y_1,Y_2}(\omega)$ is bosonic density of states \eref{eq:dos}, and $C_{AB}$ is an anomalous contribution defined by \eref{eq:K2anom}.

The cases of $|\rmi\omega_{n_1}+\rmi\omega_{n_3}|=|\rmi\omega_{n_2}+\rmi\omega_{n_4}|\sim E$ or $|\rmi\omega_{n_1}+\rmi\omega_{n_4}|=|\rmi\omega_{n_2}+\rmi\omega_{n_3}|\sim E$ can be obtained from the above expressions by appropriate permutation of operators and frequencies.

\subsection{$|\rmi\omega_{n_1}+\rmi\omega_{n_2}|\sim E$, $|\rmi\omega_{n_2}+\rmi\omega_{n_3}|\sim E$, $|\rmi\omega_{n}|\sim\Omega$, $|\rmi\omega_{n}+\rmi\omega_{m}|\sim\Omega$, $\Omega\gg E$}

The last case which we consider is the case of large frequencies $|\rmi\omega_{n}|\sim\Omega$ but finite sums $|\rmi\omega_{n_1}+\rmi\omega_{n_2}|=|\rmi\omega_{n_3}+\rmi\omega_{n_4}|\sim E$ and $|\rmi\omega_{n_2}+\rmi\omega_{n_3}|=|\rmi\omega_{n_4}+\rmi\omega_{n_1}|\sim E$ ($\Omega\gg E$).  The first $1/\Omega^2$ order terms in the high frequency expansion of \eref{K4tFT} using \eref{P_int_tot_detail} are equal to
\begin{align}
 &\sumfinonetwo1234ABCD
\nonumber\\
 &-\left[\sumfinonetwo2341BCDA\right]
\nonumber\\
 &+\sumfinonetwo4321DCBA
\nonumber\\
 &-\left[\sumfinonetwo3214CBAD\right]
\nonumber\\
 &-\left[\sumfinonetwo3421CDBA\right]
\nonumber\\
 &-\left[\sumfinonetwo1243ABDC\right]
\nonumber\\
 &+\sumfinonetwo1423ADBC
\nonumber\\
 &+\sumfinonetwo3241CBDA \,.
\end{align}
The other cases can be obtained by appropriate permutation of operators and frequencies.

The order of magnitude of the terms in high frequency expansion strongly depends on the way we increase the frequencies:
\begin{enumerate}
\renewcommand{\labelenumi}{(\arabic{enumi})}
	\item for the general case of $|\rmi\omega_{n}|\sim\Omega$ and $|\rmi\omega_{n}+\rmi\omega_{m}|\sim\Omega$ ($\Omega\gg E$), we have contributions of the order $1/\Omega^4$;
	\item for the case when one Matsubara's frequency, e.g., $|\rmi\omega_{n_1}|\sim E$, is finite and all the other are large $|\rmi\omega_{n}|\sim\Omega$ and $|\rmi\omega_{n}+\rmi\omega_{m}|\sim\Omega$, we have contributions of the order $1/\Omega^3$;
	\item for the case when one or two sums of Matsubara frequencies are finite, we have contributions of the order $1/\Omega^2$ with additional spikes (zero-frequency anomalies) when these sums of frequencies are equal to zero.
\end{enumerate}

\section{Summary}

In conclusion, we have presented a general approach of derivation of spectral relations for the four-time fermionic Green's functions completed by the consideration of the zero-frequency anomalies. It is known that for the two-time Green's functions such anomalies contribute only in the bosonic functions and do not exist for the fermionic ones. Here, we have shown that zero-frequency anomalous terms are present in spectral representations for multi-time fermionic Green's functions when the sum of any two fermionic Matsubara frequencies is equal to zero.

Equation \eref{K4tFT} together with \eref{eq:4timelehman} and \eref{eq:4timeanomal} provides a spectral representation of the four-time fermionic Matsubara Green's function in terms of spectral densities \eref{SD_4t_cont}--\eref{SD_4t_cont3}. Special consideration of the processes involving the states with the same energy values (the same states or true or accidental degeneracy) is required in order to get correct spectral representations and correct expressions of anomalous nonergodic contributions that appear to be connected by equation~\eref{eq:K4irr} with the second cumulants of the Boltzmann distribution function.

An algorithm of analytic continuations for the solution of reverse engineering problem: extraction of the spectral densities from the known expressions for four-time Matsubara Green's functions is described.

In addition, it is shown that high-frequency expansions for the four-time fermionic Green's functions demonstrate a different asymptotic behavior and have a different order of magnitude from $\Omega^{-4}$ to $\Omega^{-2}$ for different directions in the frequency space.

\section*{Acknowledgements}

I am grateful to M.~Jarrell, J.~Moreno, and D.~Galanakis for renewing my interest in this problem, to J.K.~Freericks for stimulating communications, and to I.V.~Stasyuk for useful discussions. I acknowledge the partial support by the Department of Energy, Office of Basic Energy Sciences, under Grant No.\ DE-FG02-08ER46542 for the visits to Georgetown University where this work was initiated.

\ukrainianpart

\title{Спектральні властивості чотиричасових ферміонних функцій Ґріна}
\author{А.М. Швайка}
\address{Інститут фізики конденсованих систем НАН України, вул. Свєнціцького, 1, 79011 Львів, Україна}

\makeukrtitle

\begin{abstract}
	\tolerance=3000%
		Отримано в найбільш загальній формі спектральні співвідношення для чотиричасових ферміонних функцій Ґріна. Виділено аномальні внески на нульовій частоті, які раніше були відомі тільки для бозонних функцій Ґріна, та висвітлено їхній зв’язок з другими кумулянтами функції розподілу Больцмана. Приведено високочастотні розклади чотиричасових ферміонних функцій Ґріна для різних напрямків в просторі частот.
		\keywords багаточасові функції Ґріна, спектральні співвідношення, неергодичність
\end{abstract}


\begin{thebibliography}{10}
	
	\bibitem{kubo:570}
	Kubo R., J. Phys. Soc. Jpn., 1957, \textbf{12}, No.~6, 570--586;
	\bibdoi{10.1143/JPSJ.12.570}.
	
	\bibitem{bogolyubov:589}
	Bogolyubov N.N., Tyablikov S.V., Sov. Phys.--Doklady, 1959, \textbf{4},
	589--593.
	
	\bibitem{zubarev:320}
	Zubarev D.N., Sov. Phys.--Uspekhi, 1960, \textbf{3}, No.~3, 320--345;
	\bibdoi{10.1070/PU1960v003n03ABEH003275}.
	
	\bibitem{bonch-bruevich:book}
	Bonch-Bruevich V.L., Tyablikov S.V., The {G}reen Function Method in Statistical
	Mechanics, Noth-Holland Publishing Company, Amsterdam, 1962.
	
	\bibitem{wilcox:624}
	Wilcox R.M., Phys. Rev., 1968, \textbf{174}, No.~2, 624--629;
	\bibdoi{10.1103/PhysRev.174.624}.
	
	\bibitem{stevens:1307}
	Stevens K.W.H., Toombs G.A., Proc. Phys. Soc., 1965, \textbf{85}, No.~6,
	1307--1308; \bibdoi{10.1088/0370-1328/85/6/129}.
	
	\bibitem{suzuki:882}
	Suzuki M., Prog. Theor. Phys., 1970, \textbf{43}, No.~4, 882--906;
	\bibdoi{10.1143/PTP.43.882}.
	
	\bibitem{suzuki:277}
	Suzuki M., Physica, 1971, \textbf{51}, No.~2, 277--291;
	\bibdoi{10.1016/0031-8914(71)90226-6}.
	
	\bibitem{fernandez:505}
	Fernandez J.F., Gersch H.A., Proc. Phys. Soc., 1967, \textbf{91}, No.~2,
	505--506; \bibdoi{10.1088/0370-1328/91/2/131}.
	
	\bibitem{callen:505}
	Callen H., Swendsen R.H., Tahir-Kheli R., Phys. Lett. A, 1967, \textbf{25},
	No.~7, 505--506; \bibdoi{10.1016/0375-9601(67)90012-6}.
	
	\bibitem{lucas:503}
	Lucas G.L., Horwitz G., J. Phys. A: Gen. Phys., 1969, \textbf{2}, No.~5,
	503--508; \bibdoi{10.1088/0305-4470/2/5/003}.
	
	\bibitem{morita:1030}
	Morita T., Katsura S., J. Phys. C: Solid State Phys., 1969, \textbf{2}, No.~6,
	1030--1036; \bibdoi{10.1088/0022-3719/2/6/314}.
	
	\bibitem{kwok:1196}
	Kwok P.C., Schultz T.D., J. Phys. C: Solid State Phys., 1969, \textbf{2},
	No.~7, 1196--1205; \bibdoi{10.1088/0022-3719/2/7/312}.
	
	\bibitem{ramos:441}
	Ramos J.G., Gomes A.A., Nuovo Cimento, 1971, \textbf{3}, No.~2, 441--455;
	\bibdoi{10.1007/BF02813703}.
	
	\bibitem{frobrich:014410}
	Fr\"obrich P., Kuntz P.J., Phys. Rev. B, 2003, \textbf{68}, No.~1, 014410;
	\bibdoi{10.1103/PhysRevB.68.014410}.
	
	\bibitem{mancini:37}
	Mancini F., Avella A., Eur. Phys. J. B, 2003, \textbf{36}, No.~1, 37--56;
	\bibdoi{10.1140/epjb/e2003-00315-0}.
	
	\bibitem{mancini:537}
	Mancini F., Avella A., Adv. Phys., 2004, \textbf{53}, No. 5-6, 537--768;
	\bibdoi{10.1080/00018730412331303722}.
	
	\bibitem{tyablikov:102}
	Tyablikov S.V., Pu F.C., Sov. Phys.--Solid State, 1961, \textbf{3}, 102--104.
	
	\bibitem{tanaka:388}
	Tanaka T., Moorjani K., Morita T., Phys. Rev., 1967, \textbf{155}, No.~2,
	388--392; \bibdoi{10.1103/PhysRev.155.388}.
	
	\bibitem{harper:1613}
	Harper C., Phys. Rev. B, 1972, \textbf{5}, No.~4, 1613--1618;
	\bibdoi{10.1103/PhysRevB.5.1613}.
	
	\bibitem{abrikosov:book}
	Abrikosov A.A., Gorkov L.P., Dzyaloshinski I.E., {Methods of quantum field
		theory in statistical physics}, Dover Publications Inc., New York, 1975.
	
	\bibitem{yue:4578}
	Yue J.T., Doniach S., Phys. Rev. B, 1973, \textbf{8}, No.~10, 4578--4584;
	\bibdoi{10.1103/PhysRevB.8.4578}.
	
	\bibitem{nozieres:3099}
	Nozi\`eres P., Abrahams E., Phys. Rev. B, 1974, \textbf{10}, No.~8, 3099--3112;
	\bibdoi{10.1103/PhysRevB.10.3099}.
	
	\bibitem{shastry:1068}
	Shastry B.S., Shraiman B.I., Phys. Rev. Lett., 1990, \textbf{65}, No.~8,
	1068--1071; \bibdoi{10.1103/PhysRevLett.65.1068}.
	
	\bibitem{devereaux:175}
	Devereaux T.P., Hackl R., Rev. Mod. Phys., 2007, \textbf{79}, No.~1, 175--233;
	\bibdoi{10.1103/RevModPhys.79.175}.
	
	\bibitem{ament:705}
	Ament L.J.P., van Veenendaal M., Devereaux T.P., Hill J.P., van~den Brink J.,
	Rev. Mod. Phys., 2011, \textbf{83}, No.~2, 705--767;
	\bibdoi{10.1103/RevModPhys.83.705}.
	
	\bibitem{shvaika:137402}
	Shvaika A.M., Vorobyov O., Freericks J.K., Devereaux T.P., Phys. Rev. Lett.,
	2004, \textbf{93}, No.~13, 137402; \\\bibdoi{10.1103/PhysRevLett.93.137402}.
	
	\bibitem{shvaika:045120}
	Shvaika A.M., Vorobyov O., Freericks J.K., Devereaux T.P., Phys. Rev. B, 2005,
	\textbf{71}, No.~4, 045120; \\\bibdoi{10.1103/PhysRevB.71.045120}.
	
	\bibitem{pakhira:125103}
	Pakhira N., Freericks J.K., Shvaika A.M., Phys. Rev. B, 2012, \textbf{86},
	No.~12, 125103; \bibdoi{10.1103/PhysRevB.86.125103}.
	
	\bibitem{bonch-bruevich:529}
	Bonch-Bruevich V.L., Sov. Phys.--Doklady, 1959, \textbf{129}, No.~3, 529--532.
	
	\bibitem{shvaika:447}
	Shvaika A.M., Condens. Matter Phys., 2006, \textbf{9}, No.~3, 447--458;
	\bibdoi{10.5488/CMP.9.3.447}.
	
	\bibitem{ursell:685}
	Ursell H.D., Math. Proc. Camb. Philos. Soc., 1927, \textbf{23}, No.~06,
	685--697; \bibdoi{10.1017/S0305004100011191}.
	
	\bibitem{fisher:199}
	Fisher R.A., Proc. London Math. Soc., 1930, \textbf{s2-30}, No.~1, 199--238;
	\bibdoi{10.1112/plms/s2-30.1.199}.
	
	\bibitem{kubo:1100}
	Kubo R., J. Phys. Soc. Jpn., 1962, \textbf{17}, No.~7, 1100--1120;
	\bibdoi{10.1143/JPSJ.17.1100}.
	
	\bibitem{brandt:365}
	Brandt U., Mielsch C., Z. Phys. B: Condens. Matter, 1989, \textbf{75}, No.~3,
	365--370; \bibdoi{10.1007/BF01321824}.
	
	\bibitem{metzner:324}
	Metzner W., Vollhardt D., Phys. Rev. Lett., 1989, \textbf{62}, No.~3, 324--327;
	\bibdoi{10.1103/PhysRevLett.62.324}.
	
	\bibitem{georges:13}
	Georges A., Kotliar G., Krauth W., Rozenberg M.J., Rev. Mod. Phys., 1996,
	\textbf{68}, No.~1, 13--125;\\ \bibdoi{10.1103/RevModPhys.68.13}.
	
	\bibitem{falicov:997}
	Falicov L.M., Kimball J.C., Phys. Rev. Lett., 1969, \textbf{22}, No.~19,
	997--999; \bibdoi{10.1103/PhysRevLett.22.997}.
	
	\bibitem{freericks:1333}
	Freericks J.K., Zlati\'{c} V., Rev. Mod. Phys.,
	2003, \textbf{75}, No.~4, 1333--1382; \bibdoi{10.1103/RevModPhys.75.1333}.
	
	\bibitem{shvaika:177}
	Shvaika A.M., Physica C, 2000, \textbf{341--348}, No. Part 1, 177--178;
	\bibdoi{10.1016/S0921-4534(00)00435-4}.
	
	\bibitem{freericks:10022}
	Freericks J.K., Miller P., Phys. Rev. B, 2000, \textbf{62}, No.~15,
	10022--10032; \bibdoi{10.1103/PhysRevB.62.10022}.
	
	\bibitem{shvaika:349}
	Shvaika A.M., J. Phys. Stud., 2001, \textbf{5}, No. 3/4, 349--354;
	\url{http://physics.lnu.edu.ua/jps/2001/3/abs/a349_354.html}.
	
	\bibitem{bickers:8044}
	Bickers N.E., White S.R., Phys. Rev. B, 1991, \textbf{43}, No.~10, 8044--8064;
	\bibdoi{10.1103/PhysRevB.43.8044}.
	
	\bibitem{janis:11345}
	Jani\v{s} V., Phys. Rev. B, 1999, \textbf{60},
	No.~16, 11345--11360; \bibdoi{10.1103/PhysRevB.60.11345}.
	
	\bibitem{hubbard:82}
	Hubbard J., Proc. Roy. Soc. London A, 1967, \textbf{296}, No. 1444, 82--99;
	\bibdoi{10.1098/rspa.1967.0007}.
	
	\bibitem{slobodyan:616}
	Slobodyan P.M., Stasyuk I.V., Theor. Math. Phys., 1974, \textbf{19}, No.~3,
	616--620; \bibdoi{10.1007/BF01035575}.
	
	\bibitem{grewe:4420}
	Grewe N., Keiter H., Phys. Rev. B, 1981, \textbf{24}, No.~8, 4420--4444;
	\bibdoi{10.1103/PhysRevB.24.4420}.
	
	\bibitem{metzner:8549}
	Metzner W., Phys. Rev. B, 1991, \textbf{43}, No.~10, 8549--8563;
	\bibdoi{10.1103/PhysRevB.43.8549}.
	
	\bibitem{izyumov:15697}
	Izyumov Y.A., Letfulov B.M., Shipitsyn E.V., Bartkowiak M., Chao K.A., Phys.
	Rev. B, 1992, \textbf{46}, No.~24, 15697--15711;
	\bibdoi{10.1103/PhysRevB.46.15697}.
	
	\bibitem{brinckmann:187}
	Brinckmann J., EPL, 1994, \textbf{28}, No.~3, 187--192;
	\bibdoi{10.1209/0295-5075/28/3/006}.
	
	\bibitem{sarry:958}
	Sarry M.F., Sov. Phys.--Uspekhi, 1991, \textbf{34}, No.~11, 958--979;
	\bibdoi{10.1070/PU1991v034n11ABEH002482}.
	
\end{thebibliography}
\end{document}